\documentclass[
reprint,
]{revtex4-2}

\usepackage{float}
\usepackage{graphicx}
\usepackage{dcolumn}
\usepackage{bm}

\begin{document}

\preprint{APS/123-QED}

\title{$^{23}$Na-NMR study of the one-dimensional superoxide spin-chain 
compound NaO$_2$}

\author{Takayuki Goto}
 \email{gotoo-t@sophia.ac.jp}
\affiliation{Physics Division, Faculty of Science and Technology, Sophia University, Chiyodaku 102-8554, Japan}

\author{Mizuki Miyajima}
\affiliation{Department of Physics, Okayama University, Okayama 700-8530, Japan}%

\author{Takashi Kambe}
\affiliation{Department of Physics, Okayama University, Okayama 700-8530, Japan}%

\date{\today}

\begin{abstract}
We report $^{23}$Na-NMR study of the alkali superoxide NaO$_2$,
a candidate one-dimensional quantum spin system.  
The Knight shift, linewidth, and spin-lattice relaxation rate
$1/T_1$ were investigated down to 0.3 K in magnetic fields up to 16 T.
Below $T_{\rm S3} =$ 34 K, $1/T_1$ exhibits thermally activated behavior,
indicating the opening of a spin gap of $\Delta(10.1 {\rm T}) \simeq$ 38.0 K 
and $\Delta(3.55 {\rm T}) \simeq$ 46 K, 
The hyperfine coupling constant shows a sudden reduction across
the structural transition at $T_{\rm S2} \simeq$ 195 K,
reflecting the development the pronounced one-dimensionality in the elctron structure.
These results provide microscopic evidence that NaO$_2$ is a rare 
$\pi$-orbital-based one-dimensional spin chain that exhibits a spin-gapped ground state,
consistent with a spin-Peierls-like instability.

\end{abstract}


\maketitle

\section{\label{sec:level1}INTRODUCTION}
Low-dimensional quantum spin systems have recently attracted considerable interest. 
Quantum fluctuations together with reduced dimensionality suppress 
long-range magnetic order and enhance the possibility of exotic quantum phases, 
such as spin liquids.\cite{Kanoda_review,Vasiliev2018_review,Giamarchi,Berthier2001}
In this paper, we investigate the alkali superoxide NaO$_2$, a candidate one-dimensional quantum
spin system.
Previous study by Miyajima {\em et al.} has shown that the compound undergoes a three-step
structural phase transformation from Phase I to Phase IV.
Upon cooling from room temperature, the system exhibits structural phase transition 
at $T_{\rm S1} =$240 K from Phase I ({\it space group} ${\rm Fm\bar{3}m}$) to Phase II (${\rm Pa\bar{3}}$), 
and at $T_{\rm S2} \simeq $195 K to 
Phase III (${\rm Pnnm}$) with an orthorhombic structure, and finally at $T_{\rm S3} =$34 K to 
Phase IV, for which the space group has not been clarified yet. 
In the Phase IV, the drastic reduction in the magnetic susceptibility $\chi$ was observed, 
indicating
the energy gap $\Delta =$51.2 K in the spin excitation spectrum\cite{Miyajima_Kambe}.
Existence of the gap is also confirmed by inelastic neutron scattering and $\mu$SR experiments.

Unlike the other alkali metal superoxides, such as KO$_2$\cite{Smith_KO2,KO2_Bosch1980}, 
RbO$_2$\cite{Fahmi_muSR} and 
CsO$_2$\cite{Klanjsek,Ewings_Cs,Miyajima_Cs,Jeschke}, this system 
does not show any long range magnetic order down to 0.3 K, as proved by $\mu$SR measurement.
By the detailed XRD analysis on the atomic position in Phase III, they have revealed that
''O$_2^-$-dumbbells'' are aligned along $c$-axis, making a good one-dimensionality in this system.
Although this strongly indicates that the gap originates 
from the spin-Peierls transition\cite{Hase1993,Uchinokura_CuGeO3,Itoh_CuGeO3_No_dimering}, 
no direct evidence of atomic dimerization was found so far. 
In this paper, $^{23}$Na NMR measurements provide microscopic insight into 
the ground state of this compound and the one-dimensional nature of the magnetic interactions.

\begin{figure}[h]
\centering
\includegraphics[width=0.6\columnwidth,bb=40 355 430 760,clip]{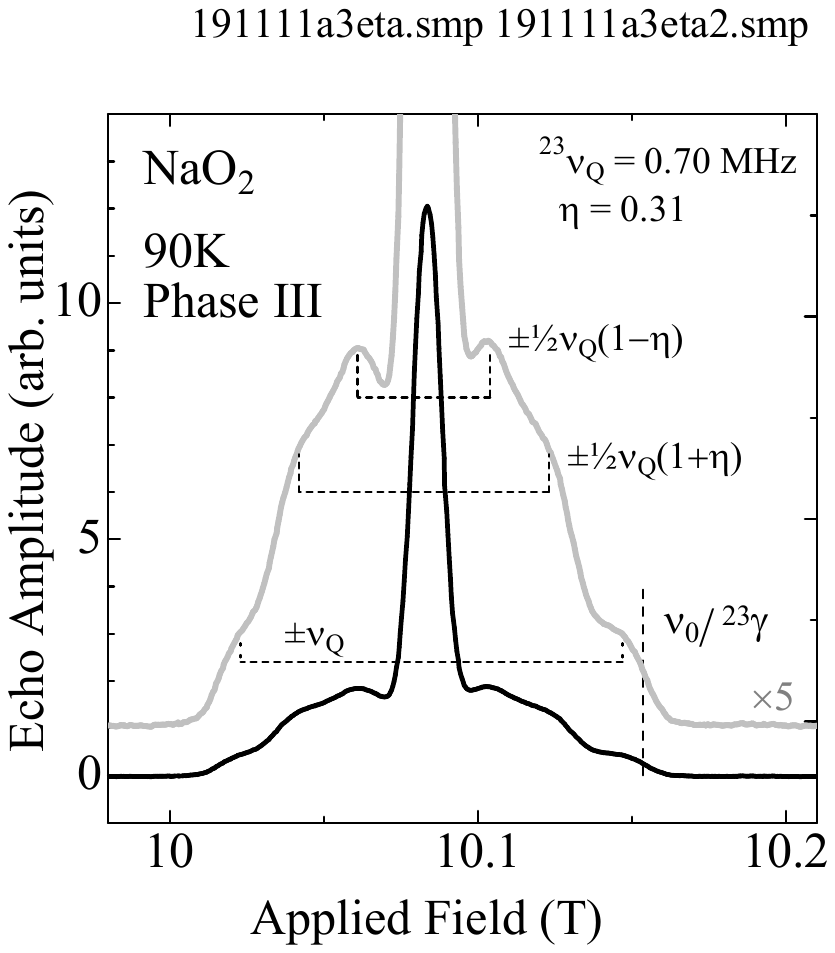}
\caption{\label{fig1}Typical $^{23}$Na-NMR spectra of powder sample measured at 90 K (Phase III).
Dashed vertical line shows the zero-shift position $\nu_0/^{23}\gamma$, 
where $\nu_0$ is resonance frequency, and $^{23}\gamma = 11.262$ MHz/T the nuclear gyromagnetic ratio for $^{23}$Na.
The enlarged spectrum (grey curve) highlights the singular points
associated with the quadrupole splitting of the $I = 3/2$ nucleus, shown by dotted lines.
From the spacing of these features, quadrupole frequency
$^{23}\nu_{\rm Q} =$ 0.7 MHz and the asymmetry parameter
$\eta =$0.31 were obtained.
}
\end{figure}

\begin{figure}[h]
\centering
\begin{minipage}{0.43\columnwidth} 
  \centering
  \includegraphics[width=1.1\linewidth,bb=43 410 458 810,clip]{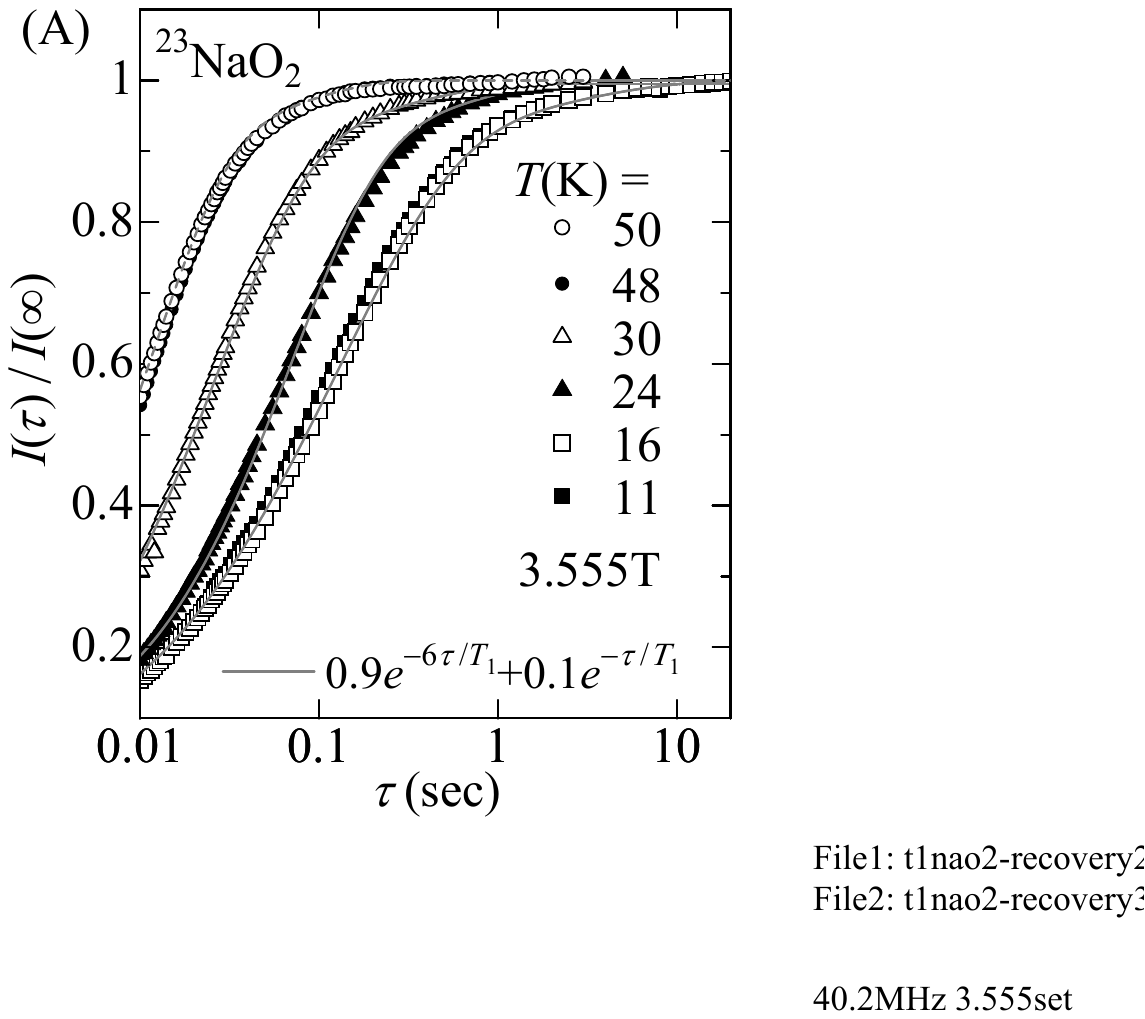}
\end{minipage}
\begin{minipage}{0.43\columnwidth}
  \centering
  \includegraphics[width=1.1\linewidth,bb=43 410 458 810,clip]{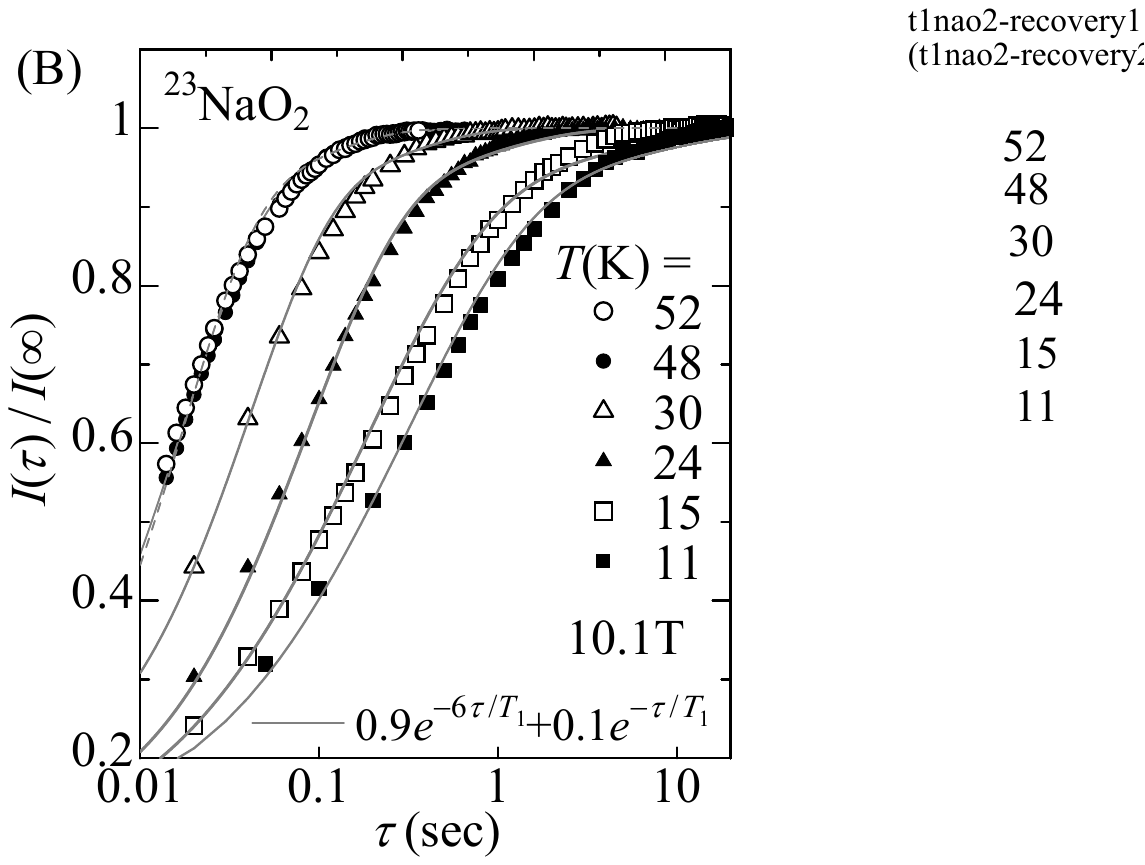}
\end{minipage}
\caption{\label{fig2}Representative recovery curves of the nuclear spin 
magnetization\cite{Takigawa2014,Morioka2026} 
under the two magnetic fields 3.554 T (A) and 10.1T (B).
The spin-lattice relaxation time $T_1$ is determined by fitting to the
standard relaxation function for the central transition for $I = 3/2$ spins 
(solid curves)\cite{Narath}.
}
\end{figure}

\section{\label{sec:level2}EXPERIMENTAL}
The powder sample of NaO$_{2}$ is prepared by the mixed solution method using liquid ammonia NH$_3$ and methyl amine CH$_3$NH$_2$ as 
described in ref. \cite{Miyajima_Kambe}.
A sample of approximately 0.1 g was sealed in an airtight Teflon cell 
in Ar-atmosphere in the glove box.
$^{23}$Na-NMR was measured in applied magnetic fields between 
3 and 16 T over the temperature range between
0.3 and 250 K.
The spectra were obtained by plotting the amplitude of spin echo against applied field, 
which was slowly ramped\cite{Watanabe_AF,Goto_Shuriken}.
Figure \ref{fig1} shows entire profile of spectrum measured at 90 K (Phase III). 
One can clearly see the singular points corresponding to the 
satellite transitions ($I_Z = \pm1/2 \leftrightarrow \pm3/2$), $90^\circ$-peaks breaking apart
due to the appreciably large value of $\eta$, which is proportional to the local orthorhombicity,
and $0^\circ$-edges\cite{Goto_Shuriken,Watanabe_AF}. 
These observations confirm the high quality of the sample.

The nuclear spin-lattice relaxation rate $1/T_1$ was determined 
by the repetition method, 
in which the nuclear spin magnetization was traced as changing the measurement interval $\tau$, until 
difference from the thermal equilibrium value became less than one percent.
\cite{Matsui_Nematic}
Figure \ref{fig2} shows typical recovery curves\cite{Takigawa2014,Morioka2026} obtained at several temperatures.
Recovery curves were analyzed by the standard formula $1-0.9e^{-6\tau/T_1}-0.1e^{-\tau/T_1}$
for central transition of nuclear spin levels $I_z = \pm 1/2$\cite{Narath}.

The distance between nearest neighboring Na and O is 2.424 \AA{} in Phase III, and 2.421 \AA{} in Phase II.
These short distances indicate that $^{23}$Na-NMR is expected to probe the magnetism 
of $S = 1/2$ spin in O$_2^-$ molecule.  We will discuss the hyperfine coupling later.

\begin{figure}[h]
\includegraphics[width=0.6\columnwidth,bb=64 330 474 738,clip]{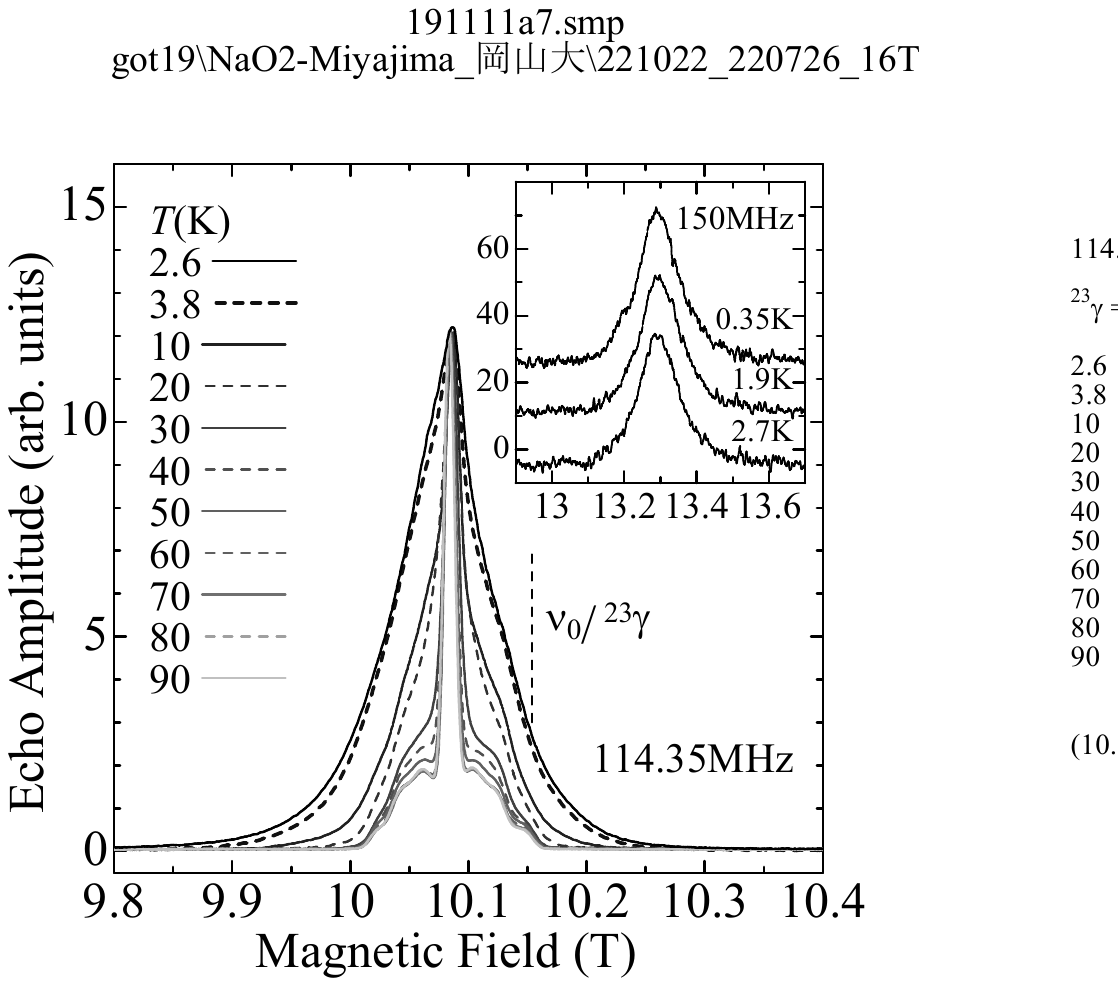}
\caption{\label{fig3}Powder $^{23}$Na-NMR spectra measured at several 
temperatures in Phase III and IV.
The dashed vertical line marks the zero-shift position of the 
$^{23}$Na nucleus. 
Inset shows the spectra obtained at low temperatures down to 0.3 K, utilizing a $^3$He-cryostat.
}
\end{figure}

\begin{figure}[h]
\includegraphics[width=0.7\columnwidth,bb=35 65 455 550,clip]{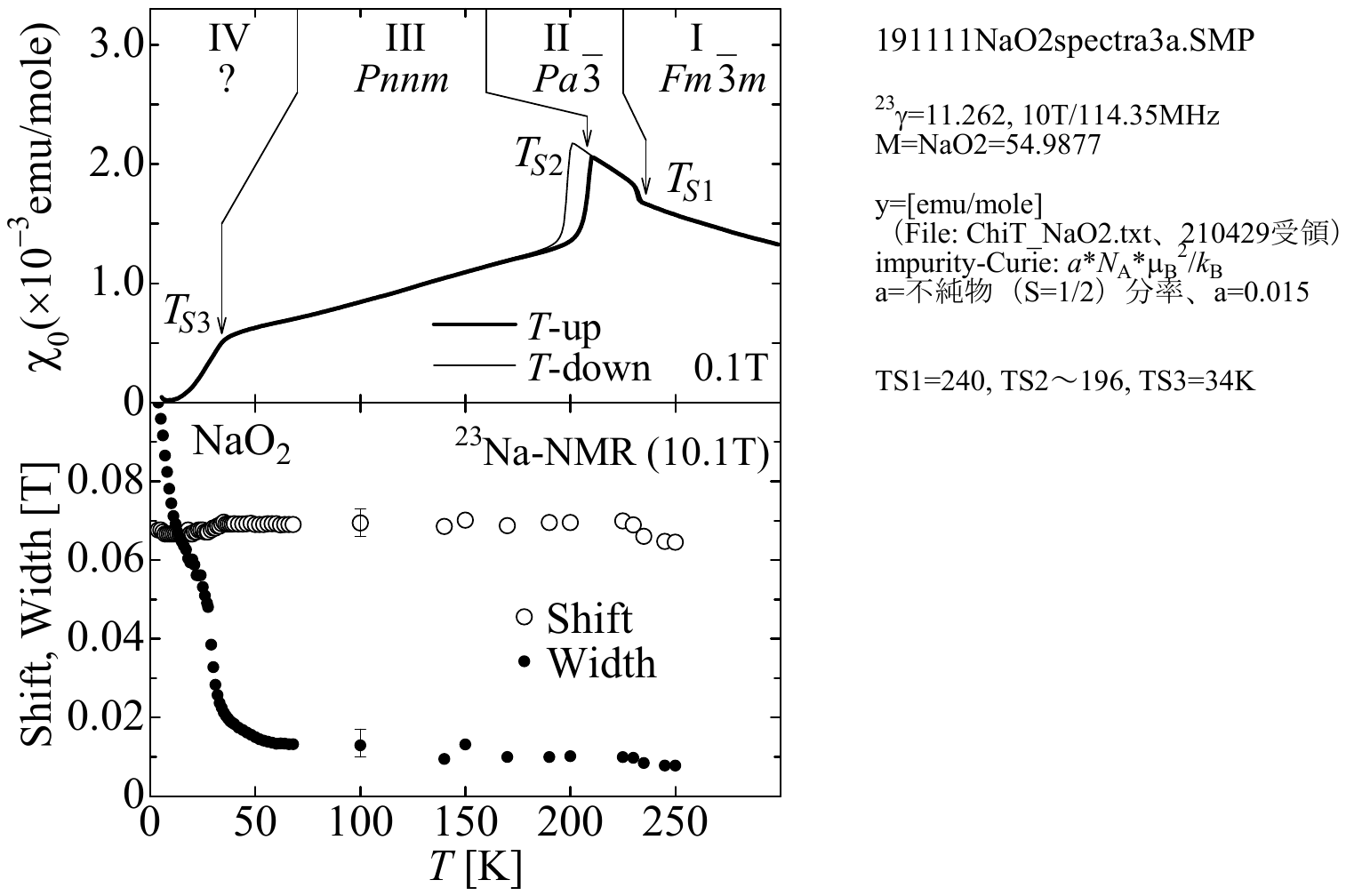}
\caption{\label{fig4}(Top) Temperature dependence of the magnetic 
susceptibility, with  
the low temperature Curie term, corresponding to 1.5 \% of free spin by mole, subtracted.\cite{Miyajima_Kambe} 
Arrows indicate the structural transition temperatures $T_{\rm s1}$, $T_{\rm s2}$ and $T_{\rm s3}$.
(Bottom) Temperature dependence of the Knight shift and the linewidth 
(FWHM) of the $^{23}$Na-NMR spectrum at 10.1 T.
}
\end{figure}

\section{\label{sec:level3}Results}
Figure \ref{fig3} shows powder patterns of entire spectra at various temperatures  
between 90 and 2.6 K in Phase III and IV.  
While the position of center peak scarcely moves,
the linewidth shows a pronounced increase as the temperature is lowered.

\begin{figure}[h]
\includegraphics[width=0.6\columnwidth,bb=33 115 455 520,clip]{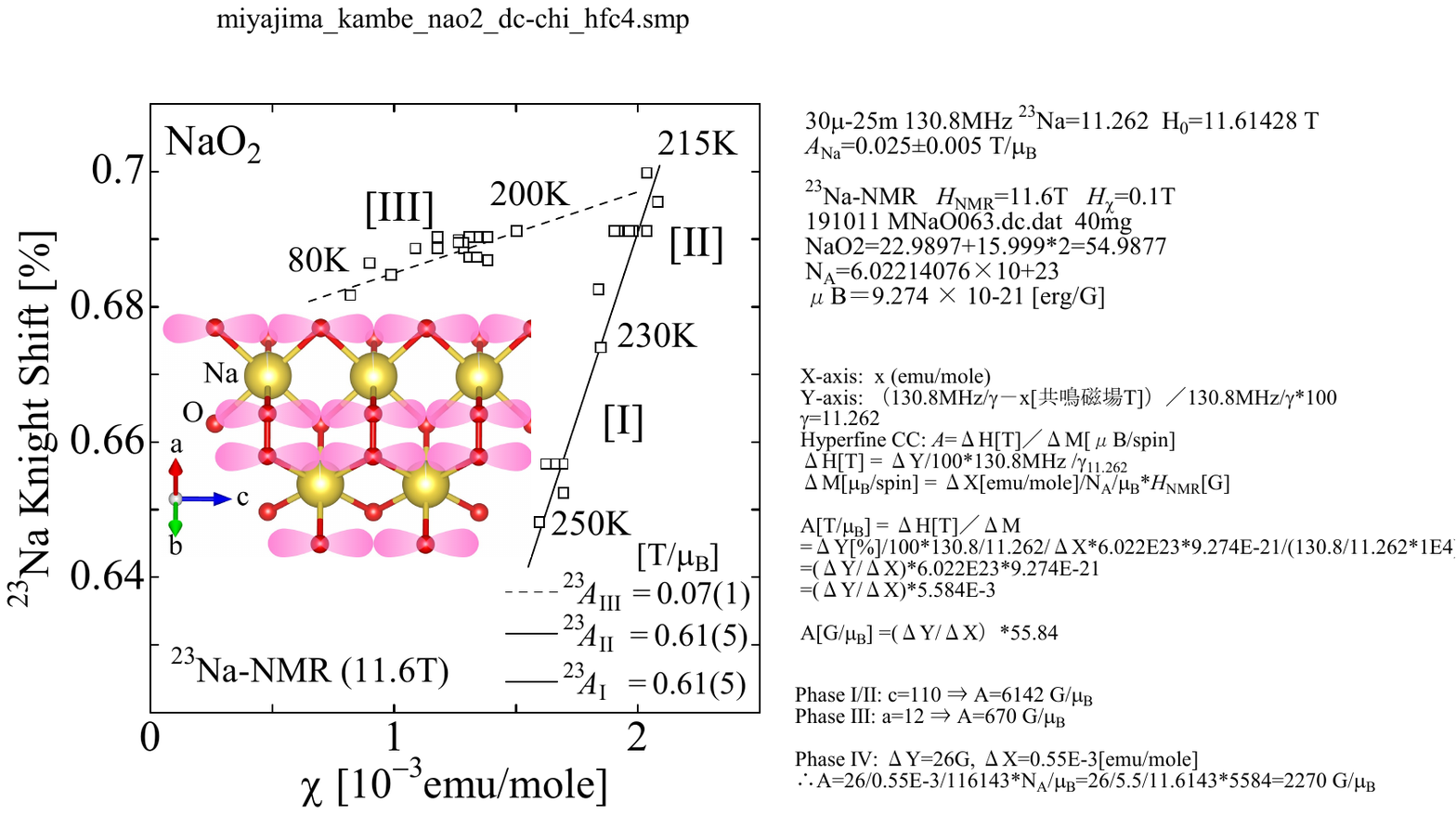}
\caption{\label{fig5}
Knight shift versus uniform susceptibility ($K$-$\chi$ plot) in Phases
I-III.  In each phase, the data exhibit a linear relationship, from which the isotropic hyperfine
coupling constant $A$ was extracted: $A_{\rm I} =$ 0.61(2),
$A_{\rm II} =$ 0.61(2) and $A_{\rm III} =$ 0.07(1) T$/\mu_{\rm B}$. 
Inset shows the schematic crystal structure in Phase III, where the O$_2$ 
molecules are aligned along $c$-axis (see text).
}
\end{figure}

The temperature dependence of the Knight shift and the linewidth is shown in Fig. \ref{fig4}.
First, the Knight shift, that is, the local susceptibility at the Na site,
exhibits only a weak temperature dependence.
With decreasing temperature from 250 K, where the system resides in Phase I,
the shift shows a slight increase in proportion to the uniform susceptibility $\chi$,  
but keeps nearly constant within Phase III.  
Below $T_{\rm S3}$, where $\chi$ exhibits a drastic reduction to zero,
the shift follows it and slightly decreases.
Proportionality between the two is shown as $K$-$\chi$ plot (Fig. \ref{fig5}),
where one can see that the isotropic hyperfine coupling constant $A$ defined as
$K = A\chi + const.$, exhibits an abrupt decrease at $T_{\rm S2}$
from 0.61(5) T$/\mu_{\rm B}$ in Phase I and II 
to 0.07(1) T$/\mu_{\rm B}$ in Phase III\cite{hyperfine_T_dep}.
This abrupt decrease in $A$ is due to the fact that the transition from Phase II 
$(Pa\bar{3})$ to III $(Pnnm)$ is the first-order structural phase transition, 
which is confirmed by $\chi$ (Fig. \ref{fig4}) and by NMR spectral profile (Fig. \ref{fig5}),
both of which exhibits a hysteresis in the thermal cycle.
This finding directly reflects the one-dimensional character in Phase III, as
discussed below.

Below $T_{\rm S3}$, both $K$ and $\chi$ exhibit a distinct decrease. 
Especially the latter goes to zero with lowering temperature.
In Phase IV, there is a significant non-linearity between $M$ and 
the applied field\cite{Miyajima_Kambe}, indicating that the comparison
of $\chi$ and the NMR shift measured under different magnetic fields 
loses its meaning.  
We, therefore, did not try to perform 
$K$-$\chi$ plot in this region.
However, we can safely say at least that the observed decrease in $K$ 
must reflects that in $\chi$.

Next, we proceed to the temperature dependence of linewidth, which monotonically
increases from 250 K with decreasing temperature.
It abruptly starts to increase at around the border between Phase III and IV,
exhibiting a kind of shoulder at around 20 K,
and steeply increases again down to 2.6 K, where the FWHM reaches 0.1 T.
At still lower temperature, it saturates down to 0.35 K.

\begin{figure}[h]
\includegraphics[width=0.6\columnwidth,bb=60 395 460 800,clip]{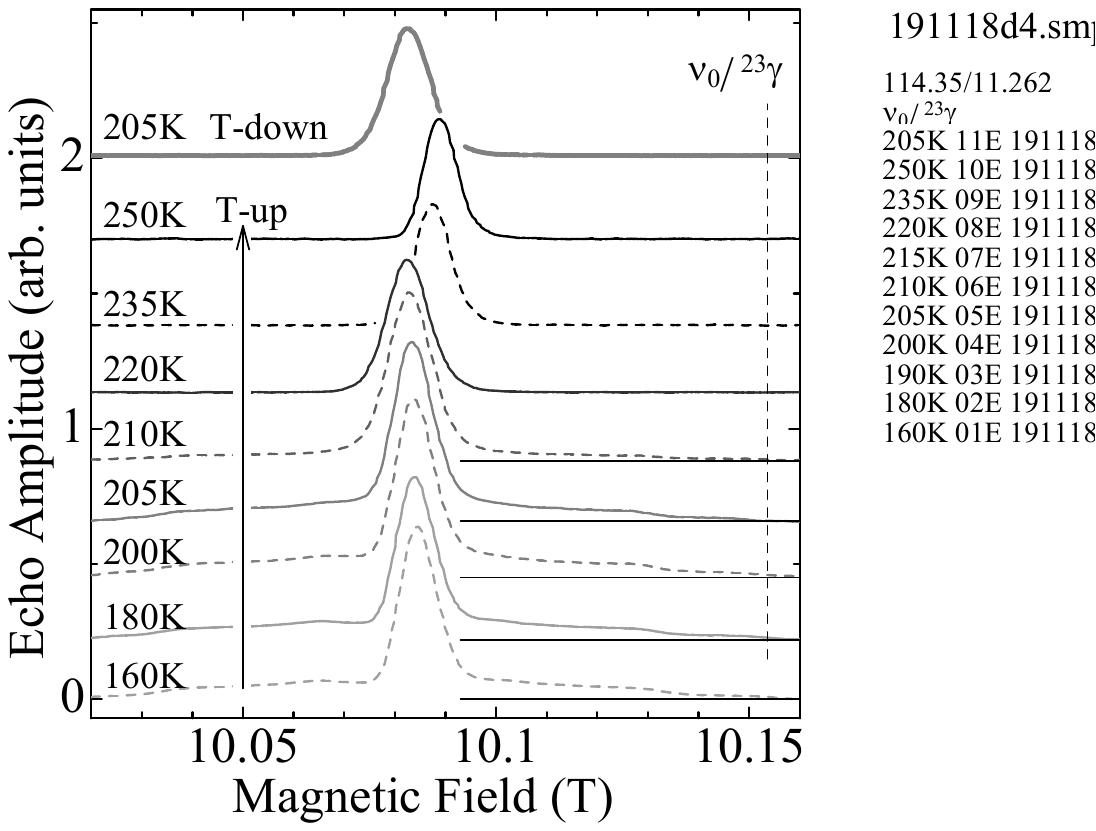}
\caption{\label{fig6}
$^{23}$Na NMR spectra measured in Phase I - III.
In order to investigate the existence of hysteresis, 
the temperature was raised from 
160 K (Phase III, bottom) to 250 K (Phase I, second from top),  
and then rapidly cooled to 205 K (Phase III, top).
The vertical solid arrow indicates the measurement sequence.
The horizontal baselines under each spectrum emphasize the spectral tails observed below 210 K.
The disappearance of the $eqQ$-derived spectral tails at 220 K signals the transition
from the orthorhombic Phase III to the cubic Phase II across $T_{\rm S2}$.
The top spectrum at 205 K shows a tail-less profile, which is characteristic to
Phase I/II.  
Note the existence of significant $eqQ$-tails in the 205 K-data measured
during the heating process.  This difference clearly demonstrates the existence of
hysteresis, and hence the first-order structural phase transition.
}
\end{figure}

The spectral profile shown in Fig. \ref{fig3} indicates that the prominent broadening 
below 40 K (\ref{fig3}) is magnetic origin rather than the electric quadrupole ($eqQ$) interaction, 
That is to say, if the latter was the dominant contribution to the linewidth,
significant broadening should appear only in the tail part but in the 
central peak\cite{PowderPattern}, because
it contributes to the linewidth of center peak and 
satellite tails through the 2$^{nd}$ and the 1$^{st}$ perturbation, respectively.  
In fact, both the central peak and tails are broadened equally 
below 40 K, indicating that the magnetic inhomogeneity is evidently dominant.
Furthermore, one can immediately conclude that its origin is some impurity part 
rather than the intrinsic part of NaO$_2$ sample.
This is because the observed temperature dependence of linewidth does 
not trace that of $\chi$, below 40 K, 
where the linewidth starts steap increasing while $\chi$ goes down to zero.
The short distance between the Na nucleus and O$_2^-$ ion stated above must
bring rather a large contribution to the linewidth through the anisotropic part of hyperfine
interaction, including the classical dipole-dipole interaction;
here note that it does not contribute to the Knight shift.
The observed apparent discrepancy between $\chi$ and the linewidth below $T_{\rm S3}$
immediately leads to the conclusion that the linewidth is contributed from the exrinsic part of sample,
that is, from impurities, which in fact are observed as the Curie term in $\chi$.\cite{Miyajima_Kambe} 
The existence of this huge broadening prevented us to investigate a change in the eqq-interaction
below $T_{\rm S3}$.  It might give an information of unfound structural change such as dimering
\cite{Hase1993,Uchinokura_CuGeO3,Berthier2001,Horvatic1998}.

\begin{figure}[h]
\includegraphics[width=0.6\columnwidth,bb=55 65 455 475,clip]{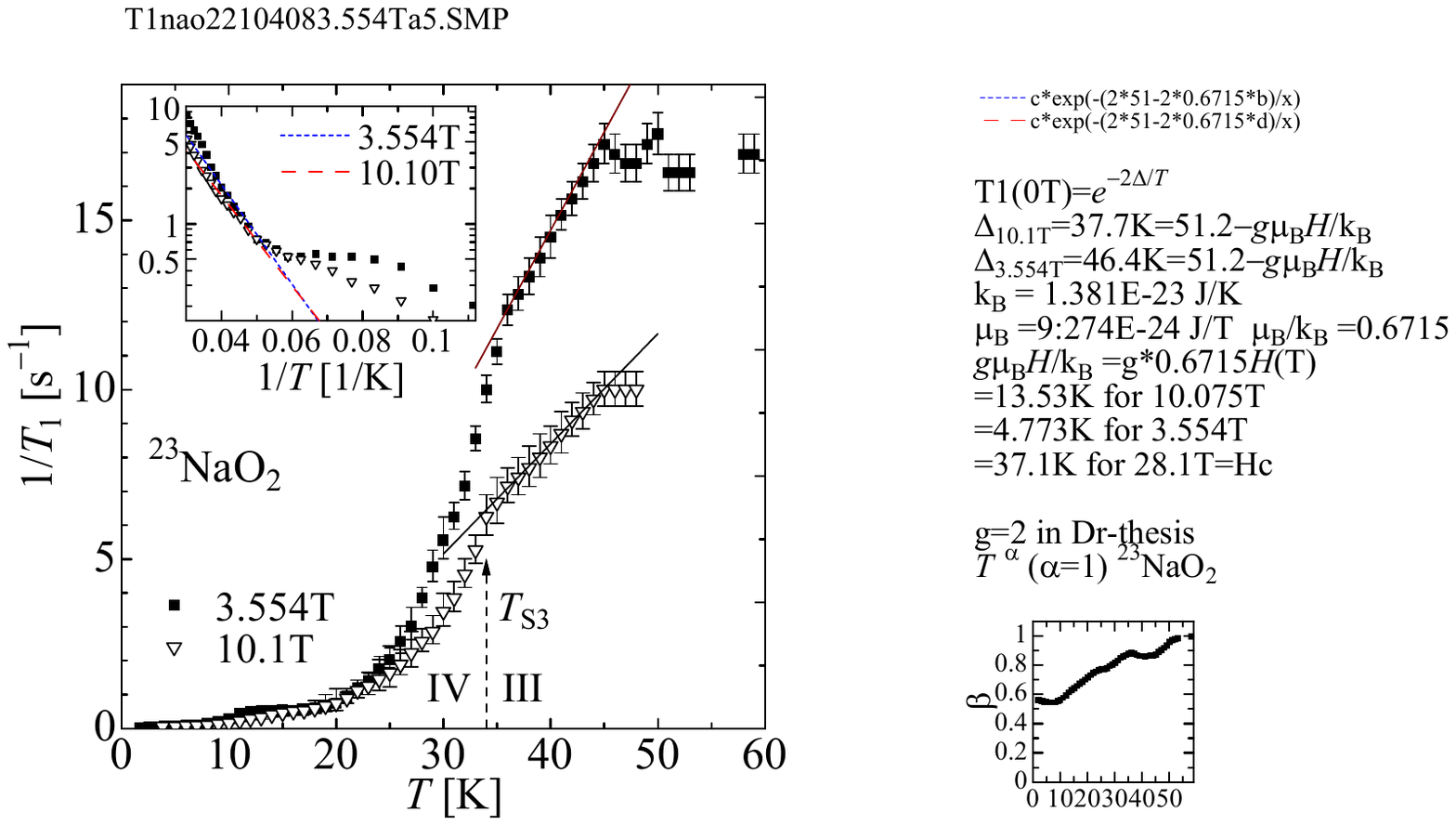}
\caption{\label{fig7}Temperature dependence of the spin-lattice
relaxation rate $1/T_1$ under the magnetic fields 3.554 and 10.1T.
The vertical dashed arrow marks the transition temperature $T_{\rm S3}$,
the boundary between Phase III and IV.
The solid lines indicate the linear $T$-temperature dependence 
observed between 45 K and $T_{\rm S3}$.
Inset: Arrenius plot showing the thermal-activation-type temperature dependence 
of $1/T_1$ between $T_{\rm S3}$ and 20 K, below which it exhibits a hump like 
behavior.
The dashed lines indicate energy gap under mangetic field 
$\Delta =$ 38.0(2) K, at 10.1 T, and 46.4 K at 3.554 T.
}
\end{figure}

Next, we move on to the behavior of $1/T_1$ shown in Fig. \ref{fig7}.  
On cooling from 60 K in Phase III, $1/T_1$ starts
to decrease at 45 K and shows the linear dependence against temperature.  
This dependence was observed commonly for the two different magnetic fields 3.554 and 10.1 T, 
although the magnitude of $1/T_1$ was quite different.
At $T_{\rm S3}$, $1/T_1$ declines precipitously
and exhibits the thermal-activation-type temperature dependence. 
However, this tendency is ceased at around 20 K, below which $1/T_1$ 
form a small hump at 12 and 15 K for 3.554 and 10.1 T, respectively,
and decreases again gradually down to 2.5 K.
The height of the hump also depends on the applied magnetic field; it 
seems to be suppressed under higher fields.

The size of energy gap under finite magnetic fields, shown in 
the inset of Fig. \ref{fig7},
$\Delta = $38.0(2) K for 3.554 T and 46.4(2) K for 10.1 T
is consistent with the zero-field gap 
$\Delta(0 {\rm T}) =$ 51.2 K obtained from $\chi$.\cite{Miyajima_Kambe},
if we consider that the first-excited state above the gap has
an $S = 1$ spin, that is, the magnon-like character,
where the energy level of $S_z = +1$ 
is lowered by Zeeman effect\cite{Matsui_Nematic,Nawa_Nematic,NH4CuCl3_ESR,Nojiri_CuGeO3_ESR}
under finite magnetic fields and shrinks the gap 
as $\Delta(H)=\Delta(0)-g\mu_{\rm B}H$,
where $g=2$ is obtained from X-band ESR measurement.
\cite{Miyajima_Kambe}.

\begin{figure}[h]
\includegraphics[width=0.6\columnwidth,bb=43 130 430 540,clip]{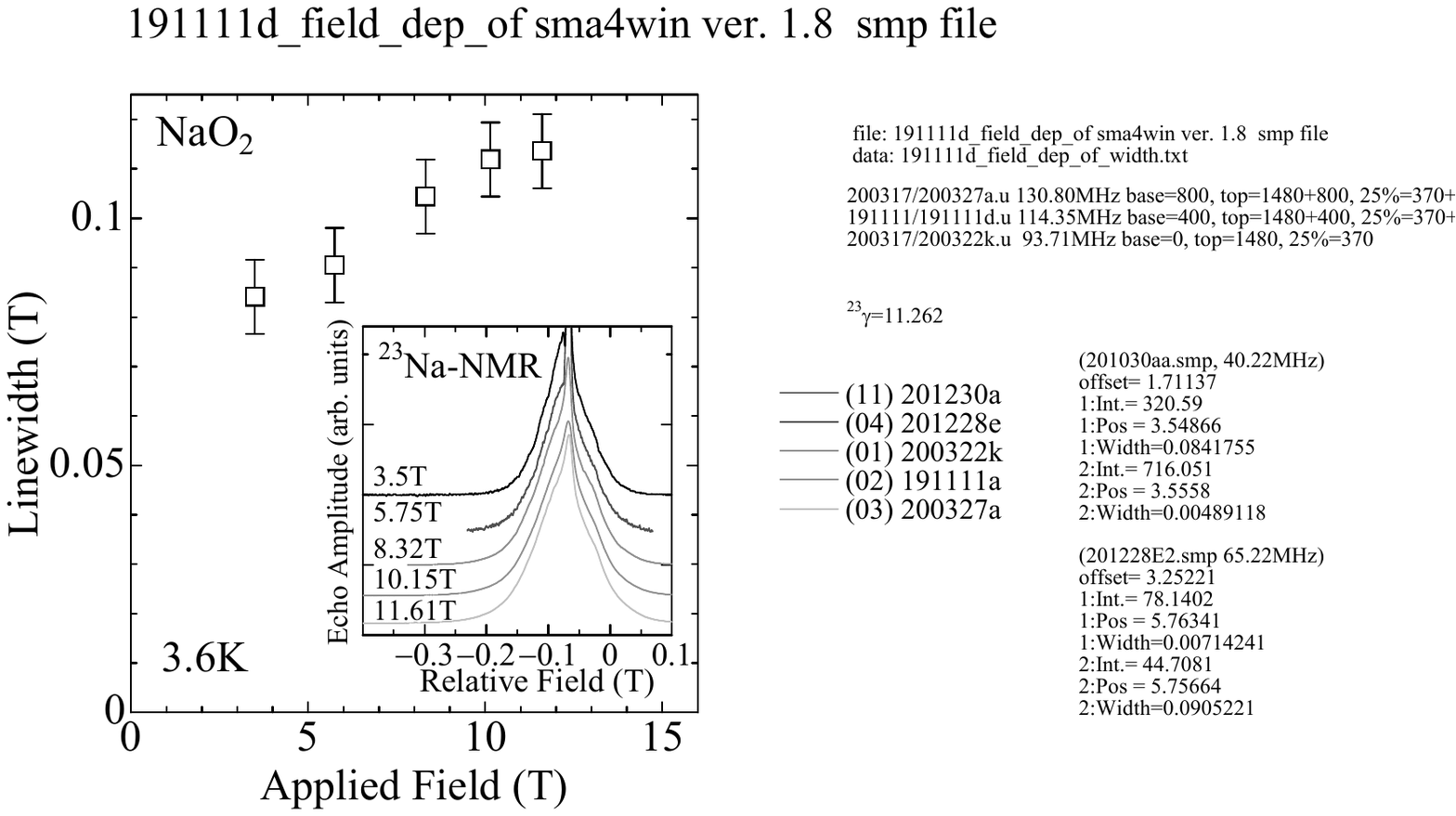}
\caption{\label{fig8}Field dependence of the $^{23}$Na NMR linewidth 
(FWHM) measured at 3.6 K.
Inset: representative spectra measured under various applied fields.
}
\end{figure}

Finally, Fig. \ref{fig8} shows the field-dependence of linewidth
at 3.6 K, where it increases monotonically with applied field $H$ up to 12 T.
This increase is also attributed to the impurities stated above.
In the previous report, the small contribution from magnetic impurities is observed
in the high-field uniform magnetization $M$.\cite{Miyajima_Kambe}

\section{\label{sec:level4}Discussion}
First, existence of an energy gap in the spin excitation spectrum
is clearly demonstrated by the observed thermal-activation-type
temperature dependence of $1/T_1$ below $T_{\rm S3}$.
The slight but finite decrease in the Knight shift below $T_{\rm S3}$ (Fig. \ref{fig4}) also supports this idea
and indicates that the ground state is non magnetic one.
In the previous paper\cite{Miyajima_Kambe}, possibility of the
spin-Peierls instability is claimed for this energy gap.  
The observed field dependence of the gap is consistent with this picture, that is,
the excited spin state of a spin-Peierls system is characterized by the breaking of 
spin-singlet dimers, leading to the creation of mobile, triplet excitations 
known as magnons.\cite{Peierls_magnon}

The presence of one-dimensionality is essential for the occurrence of a spin-Peierls transition.
The observed extremely small isotropic part of hyperfine coupling
constant within Phase III, shown in Fig. \ref{fig5}, 
provides microscopic evidence for this one-dimensionality. 
That is, the smallness in the hyperfine-coupling constant
is attributed to the lifting of the two $\pi_g$ orbitals.\cite{Miyajima_Kambe}
In Phase III, O$_2$ molecules are aligned along $c$-axis (Fig. \ref{fig5}), 
and the lobes of the occupied $\pi_g$'s with the lower energy are also directed along the axis.
This realizes the one-dimensional direct exchange coupling $J/k_{\rm B} =$ 140 K \cite{Miyajima_Kambe},
and at the same time, reduces the overlap between
the $\pi_g$ orbital and the Na$^+$ ion, and hence the hyperfine coupling
to be as small as $A_{\rm III} =$ 0.07(1) T$/\mu_{\rm B}$ in Phase III.
This is in sharp contrast  to the case of CsO$_2$, where
the lobe of $\pi_g$ is directed to Cs$^+$, resulting in the large 
hyperfine coupling constant $-$1.16 T\cite{Klanjsek}.

The power-law behavior of $1/T_1$ just above $T_{\rm S3}$ is
indicative of Tomonaga-Luttinger liquid (TLL) 
with the Luttinger parameter of $K=0.25$,
deduced from the relation $1/T_1 \propto T^{1/2K-1}$,
where the transverse component of spin fluctuation dominates $1/T_1$
\cite{Suga_Nematic,Sato_Nematic,Attractive_TLL,Klanjsek_TLL}.
Note that the effect of longitudinal component in spin fluctuation
should become visible at still lower temperature region, where however, opening
of the gap hides its effect\cite{Goto_Takao_TLL}.
The value of $K=0.25$ agrees with the theoretical value of $S = 1/2$ XXZ chain
in the low-field limit.\cite{Giamarchi}
The investigation of the field dependence of $K$ is unfortunately 
inaccessible, because of the large exchange constant $J$ and 
the saturation field.\cite{Miyajima_Kambe}
The kink of $1/T_1$ at 45 K suggests the crossover to the paramagnetic
state.  This type of precipitous change in various quantities including NMR-$1/T_1$ at
the crossover between TLL and paramagnetic state
is often reported for other 1D spin chains.\cite{Matsui_Nematic, Honda_Hagiwara,Ruegg_TLL,Willenberg_TLL}

Next, we focus on the small hump in $1/T_1$ at 13-15 K.  Although this seems to indicate
the existence of magnetic instability, no anomaly is observed in the other physical quantities,
irrespective of macroscopic or microscopic, $\chi$, $M$, $C$, $\mu$SR and the inelastic neutron scattering experiments.
We consider that this hump is brought by a slow-down of spin fluctuation in the impurity phase,
stated above.
The concentration of 1.5 \% for impurity spins, which corresponds to the mean distance of roughly 2.2 nm,
which may be too long to allow the occurrence of thermodynamical phase transition in the system.
So, these impurity spins only yields the slow down of the fluctuation below $T_{\rm S3}$, 
causing the $1/T_1$ hump by the BPP mechanism.\cite{BPP}
The similar phenomenon was reported in other one-dimensional antiferromagnet 
(CH$_3$)$_2$CHNH$_3$Cu(Cl$_x$Br$_{1-x}$)$_3$.\cite{Oosawa_IPA}

Finally, we discuss a nature of the impurities.  
Considering the existence of slow down in the spin fluctuation at the considerably high 
temperature, they are not free spins; the weak interaction via singlet dimers is expected.
This picture is consistent with the observation that the $T_1$ recovery curves consist
of a single component in the entire temerature range including the hump.
The present case differes from the impurity-doped CuGeO$_3$, where
extra NMR peaks induced by impurities were reported.
\cite{Hase_impurity_CuGeO3,Kikuchi_CuGeO3_Si,ITOH_CuGeO3_Zn}
In order to reveal its nature, measurements of $1/T_1$ and linewidth at higher field region
is necessary, which is now under progress.

\begin{acknowledgments}
This work was supported by JSPS KAKENHI Grant Number 24K08364, 25H01546 and 25K07210.

\end{acknowledgments}


\bibliography{nao2got.bib}

\end{document}